\documentclass[12pt]{amsart}
\usepackage[initials]{amsrefs}
\usepackage{url}

\usepackage{cryptologiabib}
\usepackage{graphicx}

\title[Underwater Hacker Missile Wars]{Underwater Hacker Missile Wars:
A Cryptography and Engineering Contest}

\author{Joshua Holden}
\author{Richard Layton}
\author{Laurence Merkle}
\author{Tina Hudson}

\address{Department of Mathematics, Rose-Hulman Institute of
Technology, Terre Haute, IN 47803, USA}
\address{Department of Mechanical Engineering, Rose-Hulman Institute of
Technology, Terre Haute, IN 47803, USA}
\address{Department of Computer Science and Software Engineering, Rose-Hulman 
Institute of Technology, Terre Haute, IN 47803, USA}
\address{Department of Electrical and Computer Engineering, Rose-Hulman 
Institute of Technology, Terre Haute, IN 47803, USA}

\email{holden@rose-hulman.edu}
\email{layton@rose-hulman.edu}
\email{merkle@rose-hulman.edu}
\email{hudson@rose-hulman.edu}


\begin{document}

\begin{abstract}  
    
     For a recent student conference, the authors developed a day-long
     design problem and competition suitable for engineering, mathematics and
     science undergraduates.  The competition included a cryptography
     problem, for which a workshop was run during the conference.
     This paper describes the competition, focusing on the
     cryptography problem and the workshop.  Notes from the workshop
     and code for the computer programs are made available via the
     Internet.  The results of a personal self-evaluation (PSE) are
     described.

\end{abstract}

\maketitle

\section{Introduction}

On 27 March 2004, students from across the Midwest United States
gathered at the Rose-Hulman Institute of Technology for the 2004 MUPEC
(Midwest Undergraduate Private Engineering Colleges) Conference.  This
is an annual conference sponsored by the MUPEC group, comprising the
institutions listed in Table~\ref{Table1}.  A different institution hosts the
event each year.  Participants presented papers or posters on projects
in mathematics, computer, and engineering disciplines, and also participated
in a multidisciplinary design competition.  

This paper will focus on the design competition developed by the
authors, and especially the cryptography problem which students had to
solve.  The challenge for the conference organizers is to create a
design problem suitable for students from a variety of science, mathematics
and engineering disciplines.  Our goal in designing the
competition was to create a day-long design problem suitable for
undergraduates in engineering, mathematics and science.  

\begin{table}
    
    \caption{MUPEC Member Institutions}
    \protect\label{Table1}
\begin{tabular}{|l|}
    \hline
Cedarville University \\
Indiana Institute of Technology \\
Kettering University \\
Lawrence Technological University       \\
Milwaukee School of Engineering \\
Ohio Northern University \\
Rose-Hulman Institute of Technology \\
St. Louis University    \\
Tri-State University \\
University of Evansville \\
Valparaiso University \\
\hline
\end{tabular}

\end{table}

\section{Underwater Hacker Missile Wars}

The design competition was titled ``Underwater Hacker Missile Wars''.
The idea was to design a model rocket which could be fired from
underwater and travel through the longest possible column of water.
To complicate matters, students had to solve a cryptography problem in
order to fire the rocket.  The design competition involved skills in
engineering design and analysis as well as skills in mathematics and
computer science/software engineering.  Details on the rocket design
portion of the competition may be found in~\cite{ASEE}.

The day of the conference, the thirteen student attendees were assigned to
design competition teams of either three or four members
using an automated team-assignment software package recently developed
at Rose-Hulman~\cite{teams}.  Team assignment was based on each team
having the broad mixture of multidisciplinary skills required to
successfully compete and having students work with peers from other
institutions.  

Activities related to the design competition were divided into three
broad groups: the rocket design, a cryptography workshop, and the
contest itself.  Teams started planning the design of their rockets at
10:00 a.m. Each team designated one student to be their ``hacker''.
While most of the students continued working on the problem until
noon, from 11:00 until noon the hackers (and optionally other members
of the teams) attended a workshop on cryptography conducted by Joshua
Holden and Scott Dial, a computer science major at Rose-Hulman.
(Several faculty members attending the conference also participated.)
Preparation for the contest resumed after lunch, at 1:00 p.m.  From
then until approximately 4:00 p.m., students concurrently worked on
their rocket designs, practiced breaking more of the ciphertexts
programmed into the software, and presented their papers and posters
to the judges.  (The design work was also done in a computer-equipped
classroom.)  The contest itself began at approximately 4:00.  The
conference schedule is given in Table~\ref{Table2}. (More details are
available from the web sites~\cite{MUPECurl} and~\cite{ASEEurl}, or in~\cite{ASEE}.)

\begin{table}
    \caption{Conference Itinerary}
    \protect\label{Table2}
\begin{tabular}{|l|p{3.75in}|}
    \hline
Time & Activity \\
\hline
8:00-9:00 & Registration. Coffee, juice, muffins, and fruit. Set up
posters. Last-minute team surveys. \\
\hline
9:00-9:50 & Overview of the day's activities. PSE survey. Introduction
to the design competition.  Team assignments and introduce one another. \\
\hline
9:50-10:00 & Break. \\
\hline
10:00-12:00 &Teams brainstorm, develop a strategy and begin work on
the design problem. Determine who on the 
team will be the code-breaker. At 10:50, code-breakers leave for
workshop; others continue work. \\
\hline
11:00-12:00 & Code-breaker workshop. \\
\hline
12:00-12:50 & Lunch. \\
\hline
1:00-3:30 &
Oral and poster presentations (concurrent with continuing design) at
designated times. \\
\cline{2-2}
& Design continues (concurrent with presentations), complete
nose-cones. Code-breakers practice. \\
\hline
3:00-4:00 & Nose-cones are due for oven-firing at 3:00, returned at
3:30. Snacks provided. Paint your nose-cones! 
All design and analysis documentation is finalized for judging. \\
\hline
4:00-5:00 & Competition. Underwater Hacker Missile Wars! \\
\hline
5:00-5:30 & PSE survey. Awards. \\
\hline
\end{tabular}
\end{table}

\section{The Cryptography Workshop}

The purpose of the workshop was to familiarize the students both with
the ciphers they were going to be breaking and the software they were
going to be using to assist them.  
The workshop introduced three types of ciphers: additive ciphers
(a.k.a shift ciphers or general Caesar ciphers), affine ciphers, and
two-by-two Hill ciphers (a.k.a. matrix ciphers).  For each of the
three types of cipher we took the students through a similar routine.
First, we gave a brief explanation and an example.  Then we had the
students encipher a given message by hand using a given key.  To check
their answer, we showed them how to decipher the message using custom
software, as described below.  (The workshop was held in a
computer-equipped classroom.)  After that, we talked about how to
break the cipher using frequency distributions.  We showed them how to
use the software to determine and test a probable key for the cipher.
Finally, we let the students practice breaking a set of sample ciphers
programmed into the software.  Slides from the workshop are available
on the web at~\cite{ASEEurl}, under ``Competition Materials''.

The software was written by Scott for the workshop and the
competition.  It was written in Java and distributed as a web-based
applet.  There were three functions for each cipher: construct a letter
frequency distribution (or, in the case of the Hill cipher, a digraph
frequency distribution), recover a probable key based on a (very
small) set of ciphertext-plaintext pairs, and decipher the message
based on the probable key. 

The codebreaking functions for the additive and affine ciphers, which
are letter substitution ciphers, were based on the letter frequency method.
In this method, the codebreaker prepares a ``letter frequency
distribution'' showing how often each ciphertext letter appears in the
text.  This is then compared against the known average frequency of letters in
English plaintext;  ``e'' is the most common, ``t'' is next, and
so on.  In the case of the additive cipher the key may be recovered
from the knowledge of a single plaintext-ciphertext pair:  the key
consists of the numerical value of the ciphertext letter minus the
value of the corresponding plaintext letter, modulo 26.  Thus if the
codebreaker can correctly guess the ciphertext letter corresponding to
``e'', he or she can obtain the key and decipher the message.

For the affine cipher, the numerical value of the plaintext letter is
multiplied by the first key number and added to the second key number,
modulo 26, to obtain the numerical value of the ciphertext letter.
Thus the key can be recovered from the knowledge of two
plaintext-ciphertext pairs, which sets up a system of two equations in
two unknowns which the codebreaker can (hopefully) solve.  For
example, if the codebreaker can correctly guess, from the letter
frequency distribution, the ciphertext letters corresponding to ``e''
and ``t'', he or she can obtain the key.

The Hill cipher is slightly different because it is a block
substitution cipher.  In the two-by-two case used in the workshop, each
pair of consecutive plaintext numbers is multiplied by the key matrix
modulo 26 to obtain a pair of ciphertext numbers.  Therefore
recovering the key requires the knowledge of two different
plaintext-ciphertext correspondences, each consisting of a plaintext
pair and the corresponding ciphertext pair.  In this case the
codebreaker prepares a ``digraph frequency distribution'' showing how
often each possible pair of ciphertext letters appears consecutively
in the text.  This is then compared against the average frequency of
letter pairs in English plaintext.  This is not as well known as for
letter frequencies, but it has been found that ``th'' is the most
common letter pair, followed by ``he'', and so on.  If the codebreaker
now successfully guesses the ciphertext pairs corresponding to ``th''
and ``he'', he or she can solve two matrix equations in two unknowns
and once again obtain the key.

\begin{figure}
    \centerline{\includegraphics[width=5in]{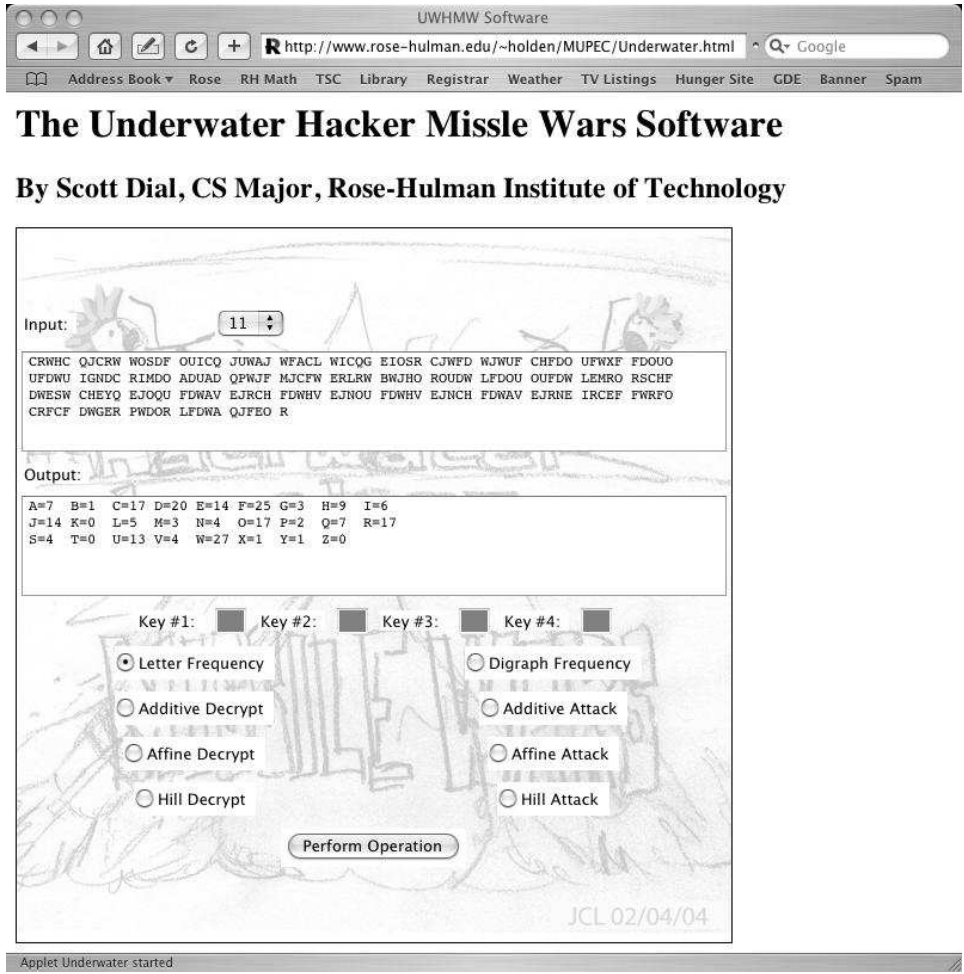}}
    \caption{Workshop software performing a letter frequency analysis}
    \protect\label{fig:frequency}
\end{figure}

\begin{figure}
    \centerline{\includegraphics[width=5in]{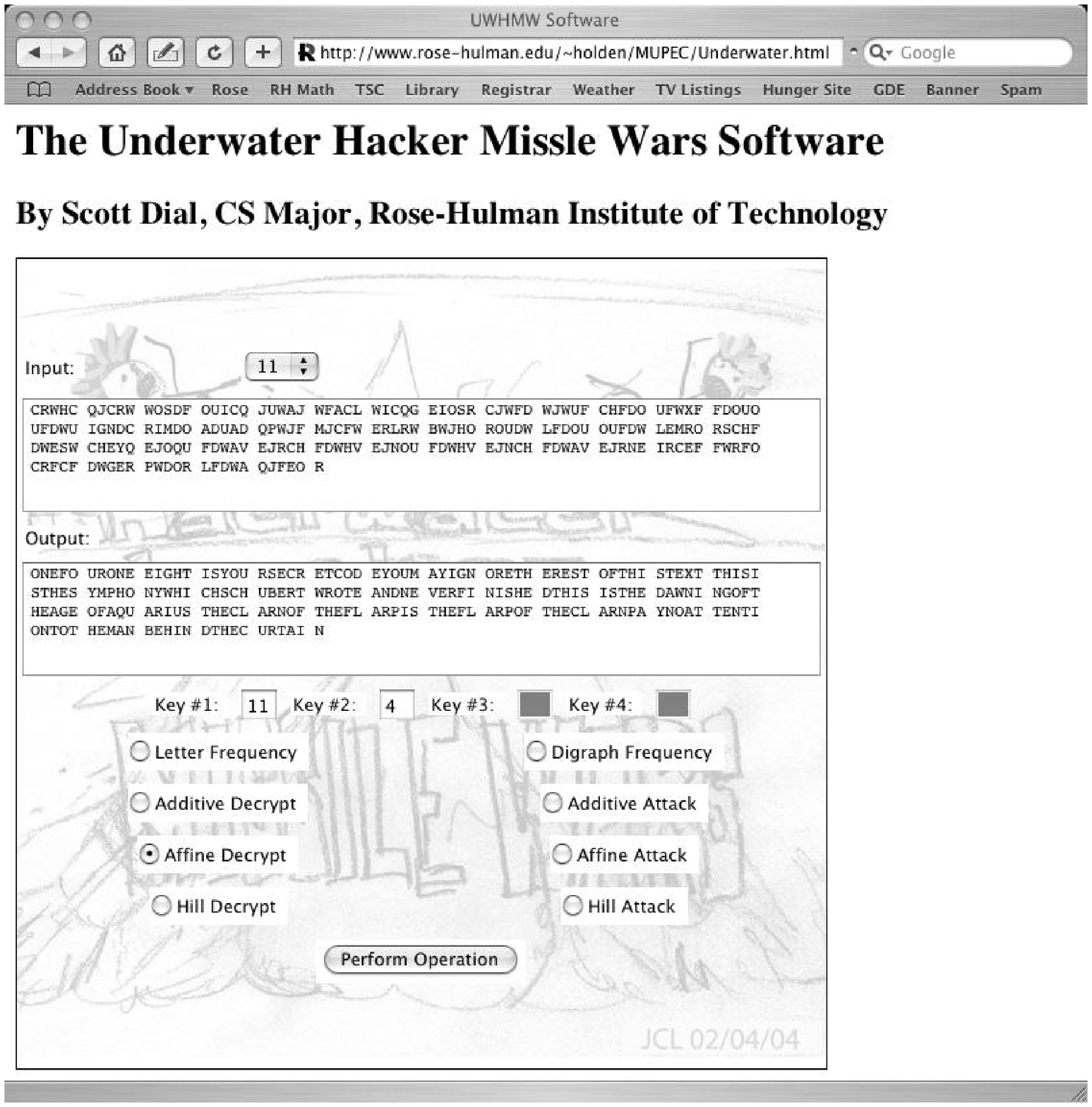}}
    \caption{Workshop software deciphering a message}
    \protect\label{fig:decrypt}
\end{figure}

The software was designed to aid this process as follows: the user
entered a ciphertext or selected one out of a sample set of
ciphertexts.  Then he or she used the software to create either a
letter frequency or digraph frequency distribution.  The user then
determined a probable key as follows: for the case of an additive
cipher, the user entered a guess for the ciphertext equivalent of
plaintext ``e''; for an affine cipher guesses for both ``e'' and ``t''
were entered; and for the two-by-two Hill cipher guesses for the pairs
``th'' and ``he'' were entered.  The ``attack'' function of the
software solved the appropriate equations and returned the
corresponding key, or a message that no such key was possible.  (The
equations were unsolvable.)  If a
key was returned, the user went on to the deciphering function and
attempted to decipher the messages.  The students were made aware that
all of the plaintext messages used in the contest were recognizable
(if not necessarily meaningful) English sentences, so that it was
immediately apparent if the key was correct.  The software, including
sample ciphertexts, is available at~\cite{ASEEurl}, under 
``Competition and Software'', and screen
shots are shown in Figures~\ref{fig:frequency}
and~\ref{fig:decrypt}.
Students were told that ciphertexts 1--10 were encrypted with an
additive cipher, ciphertexts 11--20 were encrypted with an affine
cipher, and ciphertexts 21--30 were encrypted with a Hill cipher.

Students were directed to pay special attention to the form of the
decrypted sample messages, which were constructed in exactly the same
manner as the plaintext of the messages used in the actual
competition.  Each message started with a four digit PIN (spelled
out in words), which was the only part of the message that the
students needed to know in the actual competition.  The rest of the
message consisted of several meaningless (but grammatically correct) sentences
which were chosen at random by a computer program from a list.  The
list was constructed to try to produce a large number of ``e''s,
``t''s, ``th''s, and ``he''s in order to make the frequency
distribution attack feasible with a reasonably small number of
guesses.  However, this was not completely successful for the case of 
the Hill cipher, and some of the sample texts required quite a few 
guesses.

\section{The Contest}

One ``round'' was conducted for each team; for each round an
``offensive'' and a ``defensive'' team was picked such that each 
team got exactly one offensive and one defensive opportunity.  The offensive
team's rocket was loaded into the underwater missile-launching tube.
The hackers from the offensive and defensive teams were each seated at
a laptop computer with the workshop software installed and a set of
ciphertexts loaded which they had not seen before.  All ciphertexts
used in the actual contest involved affine ciphers, and this was made
known to the contestants at the start of the contest.  Also loaded
onto the computers was control software written by Laurence Merkle
which took a round number and a four digit PIN and checked the PIN to
see if it corresponded to the ciphertext for that round.  

If the PIN was correct, the software was programmed to send a signal
to a switching module designed and built by Tina Hudson.  The
switching module was built to determine which of the two laptops had
sent the signal first.  The intended plan for the switching module was
that if the offensive team sent the signal first then the module would
produce a ``launch'' result which would connect a six-volt battery to
the ignitor of the model rocket, causing the rocket to launch.  If the
defense succeeded first the module would produce a ``no-launch''
result and the rocket would not launch.  (Due to electrical issues
this result was not conveyed directly to the rocket launcher during
the actual competition; rather an indicator light indicated which team
had succeeded first.  The electrical issues have since been resolved,
and a new launch board has been built and tested for future use.)

\begin{figure}
    \centerline{\includegraphics[width=5in]{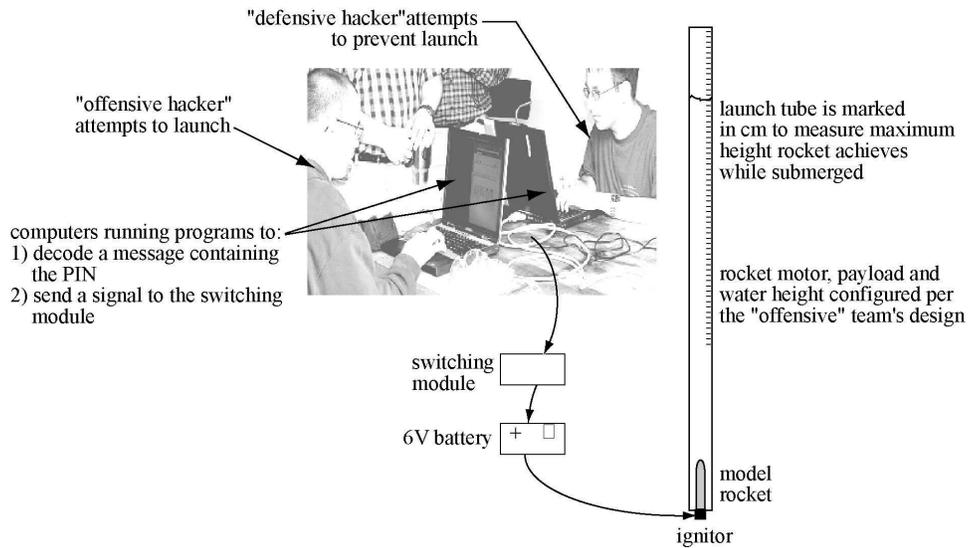}}
    \caption{Competition schematic}
    \protect\label{fig:rocket}
\end{figure}

In the actual execution, the round was started with an announcement of
the round number and the simultaneous starting of a stopwatch for each
team.  The hackers then proceeded to enter the round number into the
workshop software and get their ciphertext, which they attempted to
break.  (For fairness, both teams received the same problem.)  When
they had broken the text and obtained the four-digit number they
entered the number into the control software, which determined if it
was correct.  Each team's stopwatch was stopped when the correct code
number was recognized.  Both of the measured times were used in
scoring, so both hackers were directed to proceed until they had
broken the cipher.  The winner of the ciphertext competition was
announced, and in either case the rocket was launched so that its
performance could be used in the scoring.  Figure~\ref{fig:rocket}
presents a schematic of the competition.

Team scores for the competition as a whole were based on the
performance of the rocket (65\%), the accuracy of the team's
prediction of the performance of the rocket (12\%), the aesthetics of
the painted nose-cones of the rockets (3\%), and the code-breaking
times for both the offensive opportunity (10\%) and the defensive
opportunity (10\%).  Timing scores were calculated using the time 
measured as a fraction of the maximum time measured for any team in 
any round.  Another scoring possibility might take into account the 
``head-to-head'' nature of the competition more directly.  Also, the system 
we used did not account for the possibility that some ciphertexts 
might be more difficult to decipher than others.

\section{Conclusions}

Our goal in designing the competition was to create a day-long design
problem suitable for undergraduates in engineering, mathematics and
science disciplines.  Surveys filled out by the students support this
goal, both in general (see~\cite{ASEE}) and in the cryptography part
of the competition.  Students were originally not very confident of
their ability to break codes and ciphers and to use matrices to
encipher messages, but they indicated that they gained confidence
after attending the workshop and practicing (or watching their
teammates practice) these skills over the course of the day.  However,
the students indicated that they had lost confidence in their ability
to encrypt messages using a simple cipher.  We hypothesize that
students on average were perhaps not aware before the workshop of some
of the complexities of what could be considered a ``simple'' cipher,
and were thus overconfident.  We think the surveys indicate that on
average, some level of learning has occurred for a mixed group of
students from different disciplines and different institutions.

Quite a bit of time and effort went in to putting this competition
together.  We estimate that Prof.~Holden put in about 30 hours of work
on the design of the cryptography part of the competition and
preparing the cryptography workshop.  Prof.~Hudson spent about 20
hours for the switching module.  Prof.~Layton put in about 80 hours of
work on the design and testing of the launch-tube apparatus,
coordinating the overall competition design, and organizing the
conference.  Prof.~Merkle spent an estimated 50 hours on the control
software.  Erin Bender and Gerald Rea, mechanical engineering majors
at Rose-Hulman, put in approximately 100 hours of work on the original
analysis and simulation for the design as well as the building and
testing and redesign of the physical apparatus.  They were compensated
for this as work-study employees.  Scott Dial put in approximately 10
hours of work on the workshop software, for which he was compensated
with extra credit in Prof.~Holden's cryptography class.

However, much of this would effort would not have to be
duplicated by someone putting on a similar competition.  Complete 
plans and instructions for all aspects of the competition are or will 
be posted at~\cite{ASEEurl} and we estimate that a person or team of 
people with the appropriate expertise could reproduce the competition 
in perhaps a quarter of the time we spent.

\section*{Acknowledgements}

Our thanks to so many without whom this design problem and competition
simply would not have come together in time: our colleagues and
coworkers Ray Bland, Patsy Brackin, Gary Burgess, Pat Carlson, Mike
Fulk, and Mike McLeish, and our students Erin Bender, Scott Dial and
Gerald Rea.

\section*{Biographical Sketches}

Joshua Holden is currently an Assistant Professor in the Mathematics
Department of Rose-Hulman Institute of Technology, an undergraduate
engineering college in Indiana.  He received his Ph.D. from Brown
University in 1998 and held postdoctoral positions at the University
of Massachusetts at Amherst and Duke University.  His research
interests are in computational and algebraic number theory and in
cryptography.  His teaching interests include the use of technology in
teaching and the teaching of mathematics to computer science majors,
as well as the use of historically informed pedagogy.  His
non-mathematical interests currently include science fiction, textile
arts, and choral singing.

Tina Hudson received her Ph.D. from Georgia Institute of Technology in
2000 and is currently an Assistant Professor of Electrical Engineering
at Rose-Hulman Institute of Technology.  Her research interests
include the development of real-time neuromuscular models using
integrated circuits and MEMS devices, linear threshold circuits, and
methods to intuitively teach analog and digital integrated circuit
design and MEMS devices.

Richard Layton received his Ph.D. from the University of Washington in 1995
and is currently an Associate Professor of Mechanical Engineering at
Rose-Hulman.  His research  and teaching interests are analysis, simulation
and design of multidisciplinary engineering systems.  Prior to his academic
career, he worked for twelve years in consulting engineering, culminating as
a group head and a project manager. His non-engineering interests include
woodworking and music composition and performance for guitar and small
ensembles.  

Larry Merkle received his Ph.D. from the Air Force Institute of
Technology in 1996 and is currently an Assistant Professor of Computer
Science and Software Engineering at Rose-Hulman Institute of
Technology.  Prior to joining Rose-Hulman, he served almost 15 years
as an active duty officer in the United States Air Force.  During that
time he served as an artificial intelligence project management
officer, as chief of the Plasma Theory and Computation Center, and on
the faculty of the United States Air Force Academy.  His interests
include computer science education and the application of advanced
evolutionary computation techniques to computational science and
engineering problems.

\end{document}